\documentstyle[graphicx,natbib,epsfig,amsmath,amssymb,traditabstract,letter]{aa}

\newcommand{\nco}{\,\hbox{$N_{\rm CO}$}}
\newcommand{\nhtwo}{\,\hbox{$N_{\rm H_2}$}}

\newcommand{\fc}{\,\hbox{$f_{\rm c}$}}
\newcommand{\tex}{\,\hbox{$T_{\rm ex}$}}

\newcommand{\msun}{\,\hbox{$M_{\odot}$}}
\newcommand{\lsun}{\,\hbox{$L_{\odot}$}}
\newcommand{\kms}{\,\hbox{\hbox{km}\,\hbox{s}$^{-1}$}}

\newcommand{\htwo}{\,\hbox{$\rm{H_ 2}$}}

\newcommand{\spi}{{\it Spitzer}}

\newcommand{\hi}{\,\hbox{\ion{H}{I}}}
\newcommand{\nai}{\,\hbox{\ion{Na}{I}}}

\newcommand{\um}{\,\hbox{$\mu$m}}

\newcommand{\cooz}{\hbox{$^{12}\rm CO(1-0)$}}

\newcommand{\cozo}{\hbox{$^{12}\rm CO(0 \rightarrow 1)$}}

\newcommand{\cott}{\hbox{$^{12}\rm CO(3-2)$}}
\newcommand{\her}{\hbox{\it Herschel}}

\begin{document}
\textheight=24.8cm
\addtolength{\topmargin}{-.2cm}
\setlength{\parskip}{0.5mm plus 0.0mm minus0.0mm}
  
   \title{Cold and warm molecular gas in the outflow of 4C12.50}

   \subtitle{}

   \author{K. M. Dasyra\inst{1} 
          \and
	 F.  Combes\inst{1} 
	  }

   \institute{
             Observatoire de Paris, LERMA (CNRS:UMR8112), 61 Av. de l'Observatoire, F-75014, Paris, France \\
             }

   \date{}

    \abstract 
    { We present deep observations of the $^{12}$CO\,(1$-$0) and (3$-$2) lines in the ultra-luminous infrared and radio galaxy 4C12.50, carried out with the 30\,m telescope of the 
    Institut de Radioastronomie Millim\'etrique. Our observations reveal the cold molecular gas component of a warm molecular gas outflow that was previously known from \spi\ 
    space telescope data. The \cott\ profile indicates the presence of absorption at $-$950\kms\ from systemic velocity with a central optical depth of 0.22. Its profile is similar to that 
    of the \hi\ absorption that was seen in radio data of this source. A potential detection of the 0$\rightarrow$1 absorption enabled us to place an upper limit of 0.03 on its central 
    optical depth, and to constrain the excitation temperature of the outflowing CO gas to $\geq$65\,K assuming that the gas is thermalized. If the molecular clouds fully obscure 
    the background millimeter continuum that is emitted by the radio core, the \htwo\ column density is $\geq$1.8$\times$10$^{22}$cm$^{-2}$. The outflow then carries an 
    estimated cold \htwo\ mass of at least 4.2$\times$10$^3$\msun\ along the nuclear line of sight. This mass will be even higher when integrated over several lines of sight, 
    but if it were to exceed 3$\times$10$^9$\msun, the outflow would most likely be seen in emission. Since the ambient cold gas reservoir of 4C12.50 is 1.0$\times$10$^{10}$\msun , 
    the outflowing-to-ambient mass ratio of the warm gas (37\%) could be elevated\,with\,respect\,to\,that\,of\,the\,cold\,gas.
    }


   \titlerunning{Cold and warm molecular gas in the outflow of 4C12.50}
   \authorrunning{Dasyra \& Combes}
   
   \maketitle


\section{Introduction}
\label{sec:intro}

Through jets, winds, and radiation pressure, active galactic nuclei (AGN) can affect the collapse of their surrounding molecular gas 
and the formation of new\,stars in their\,host galaxies. While simulations attribute a considerable role to AGN feedback in shaping galaxy 
properties \citep[e.g.,][]{croton06,sijacki07,booth09,hopkins10,debuhr12},  the question remains open from an observational point of view.  
Do the AGN-driven outflows that suppress star formation occur frequently enough to affect the observed luminosity and mass functions of 
galaxies? To answer this question, we not only need to identify objects with massive, AGN-driven outflows of molecular gas, but also to take 
into account the outflow effects on the different phases of this gas. 
 
The various phases of an AGN-driven molecular gas outflow have been studied in a nearby ultraluminous infrared galaxy (ULIRG), Mrk231. 
Plateau de Bure interferometric observations showed that the outflowing CO, with a corresponding \htwo\ mass of 6$\times$10$^8$\msun\ 
and a flow rate of 700\msun\,yr$^{-1}$, is capable of suppressing  star formation \citep{feruglio10}. \citet{aalto12} discovered the outflow's dense 
($\gtrsim$10$^4$\,cm$^{-3}$) component from broad HCN(1$-$0) profile wings. On the basis of profile fitting of the 79\um\ and 199\um\ OH lines 
seen with \her\ \citep{fischer10}, the outflow contains $\ge$7$\times$10$^8$\msun\ of molecules.  In total, more than 10$^{9}$\msun\ of gas can 
be in the wind. 

Another straightforward and quantitative comparison that has yet to be performed is that between the warm and the cold \htwo\ gas mass in the 
outflow vs. in the ambient interstellar medium (ISM). A source that is suitable for this analysis is 4C12.50. It is the only source known to have a 
massive outflow of gas of few hundred Kelvin, as inferred from its mid-infrared \htwo\ rotational line profiles \citep{dasyra_combes11}. In this letter, 
we present evidence of cold molecules in its outflow, and we evaluate the relative fractions of warm and cold gas that the AGN feedback kinematically 
distorts. We adopt H$_0$=70 \kms\  Mpc$^{-1}$, $\Omega_{M}$=0.3, and $\Omega_{\Lambda}$=0.7 throughout.


\begin{figure*}
\begin{center}
\scalebox{1.02}{
\includegraphics[width=11cm]{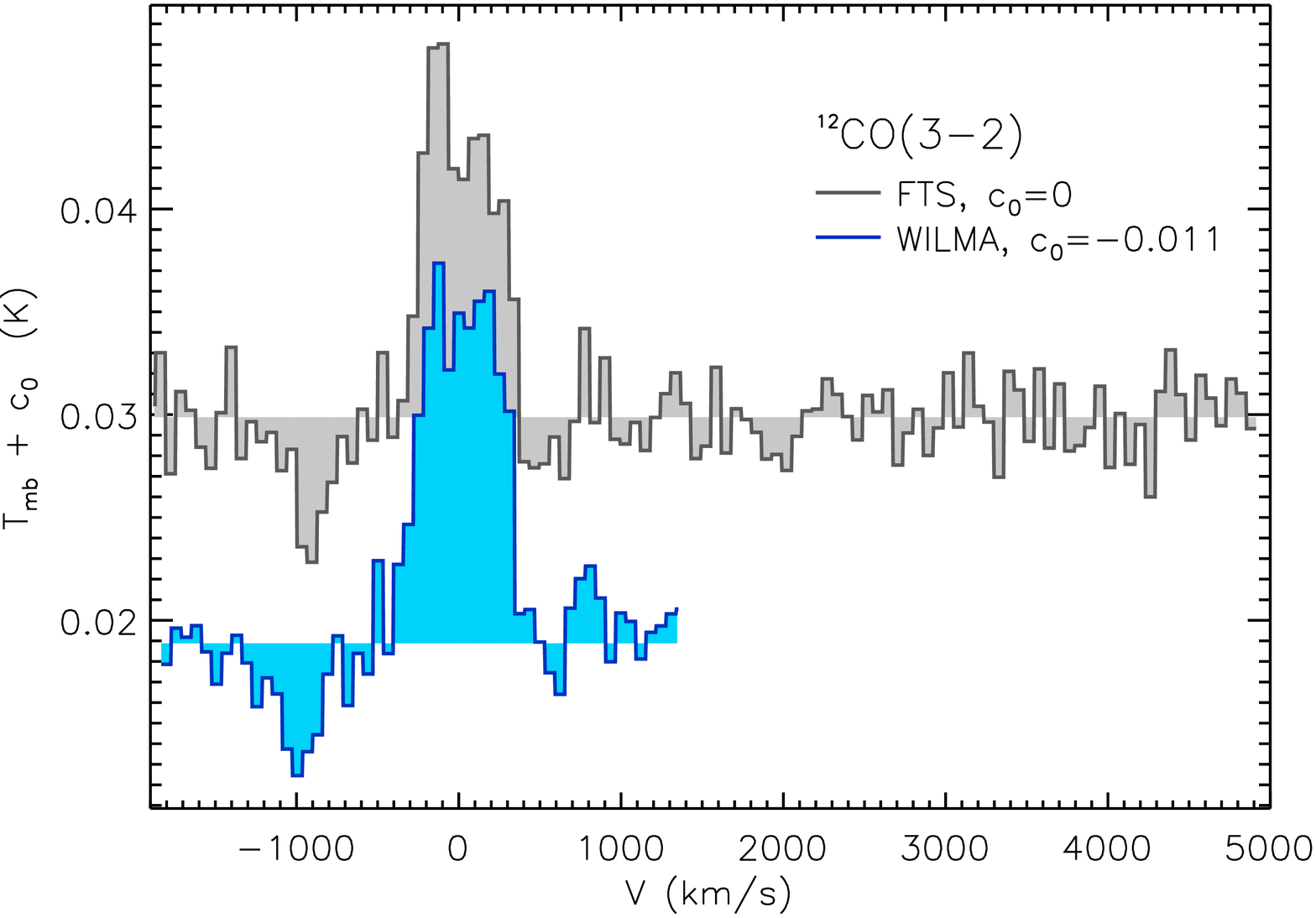} 
\includegraphics[width=7cm]{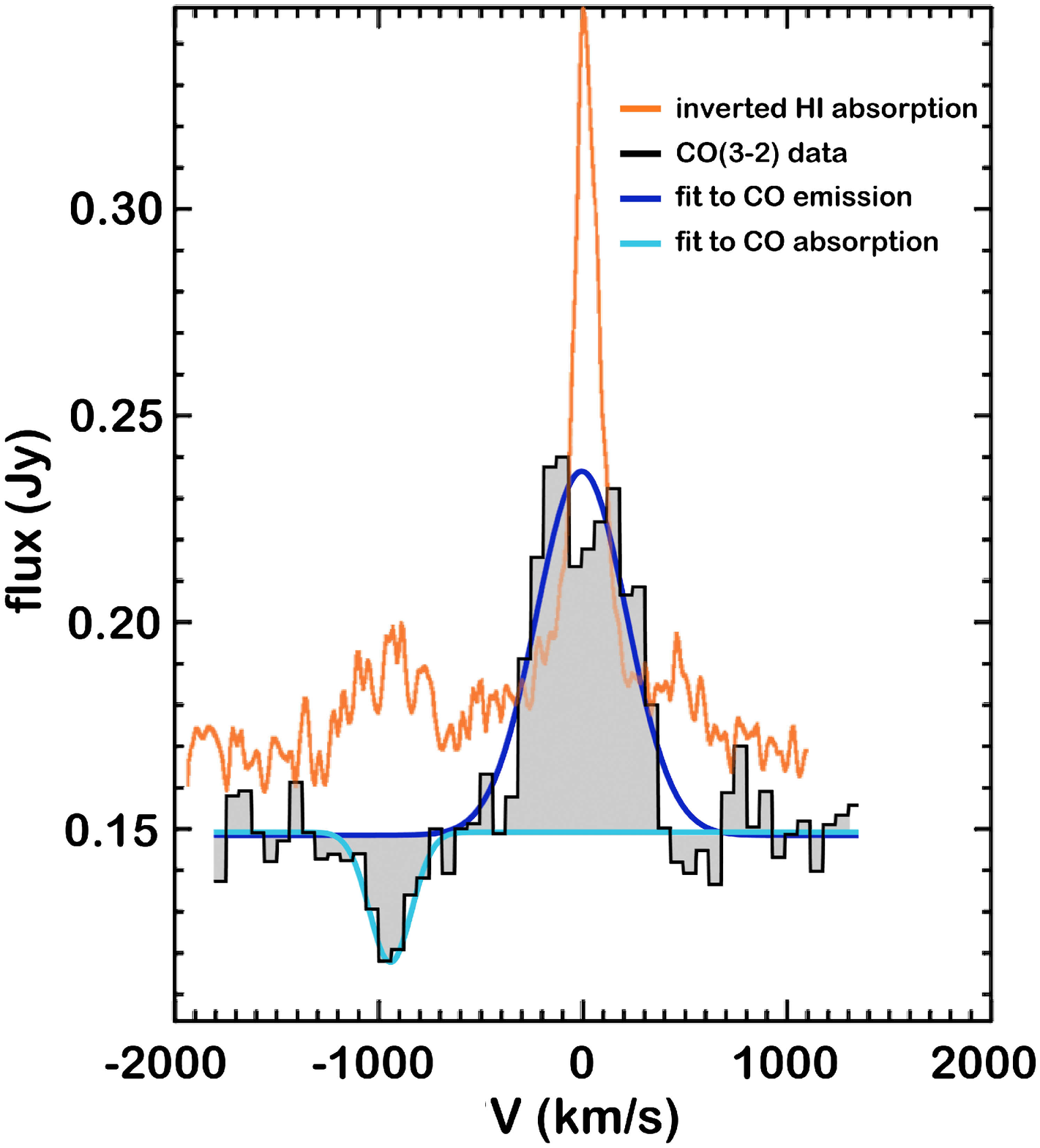} 
}
\caption{{\it Left:}  FTS and WILMA spectra of \cott\ in 4C12.50, binned into 62 \kms\ channels and shifted by a constant $c_0$, when needed. 
{\it Right:} Flux-calibrated line profile of \cott , averaged over the WILMA and the FTS data to be least affected by potential artifacts. Gaussian 
functions with parameters that most closely fit the emission component at the systemic velocity and the absorption component at $-$950\kms\ are 
shown in blue and cyan, respectively. The inverse \hi\ absorption that is seen in the radio data of 4C12.50 \citep{morganti04} is overplotted in 
orange for an arbitrary scale and continuum level.
}
\label{fig:CO32}
\end{center}
\end{figure*}

\section{Data acquisition and reduction}
\label{sec:data}
The observations were carried out with the 30\,m telescope of the Institut de Radioastronomie Millim\'etrique on 
December 21-22 2011, and January 1 2012 (as part of the program 235-11). For the \cooz\ observations, EMIR 
was tuned to 102.700\,GHz. For the \cott\ observations, the receivers were tuned to 308.086\,GHz during the first observing 
run and to 308.600\,GHz during the second observing run, to ensure that the observed line properties were unaffected by 
possible standing waves, atmospheric line residuals, or bad channels.  Parallel to the CO observations, we carried out HCN 
and HCO$^+$\,(2$-$1) observations with a common tuning of 158.427\,GHz. Both the FTS and WILMA backends, with corresponding 
resolutions of 0.2 MHz and 2MHz, were simultaneously used as backends to allow consistency checks. The inner and outer 
parts of the FTS were used, enabling us to better sample the continuum near each line. The wobbler switching mode with a 
throw of 80\arcsec\ and 0.65 sec per phase was used to rapidly sample the continuum. The telescope pointing accuracy was 
2\arcsec , and the system temperature varied from 310\,K to 580\,K at 1\,mm.

To reduce the 1\,mm data, we removed bad channels from all individual scans. The three parts of each FTS scan, 
corresponding to the output of a different spectrometer, were fitted with linear baselines and brought to the same continuum 
level, i.e., to the median value of the full scan. This step was performed to ensure that there is no continuum discontinuity 
between the three parts that could introduce artificial features into the combined spectrum. We averaged the individual scans 
of each frequency tuning and backend, discarding scans with unstable baselines. The inner and the outer FTS spectra were 
then stitched together. For each backend, the spectra of the two frequency tunings were brought to a common reference 
frequency of 308.245\,GHz and averaged. The final spectra are shown in Figure~\ref{fig:CO32}. Their total on-source integration 
time is  2.7 hours. 

\begin{figure*}
\begin{center}
\scalebox{1.01}{
\includegraphics[width=13cm]{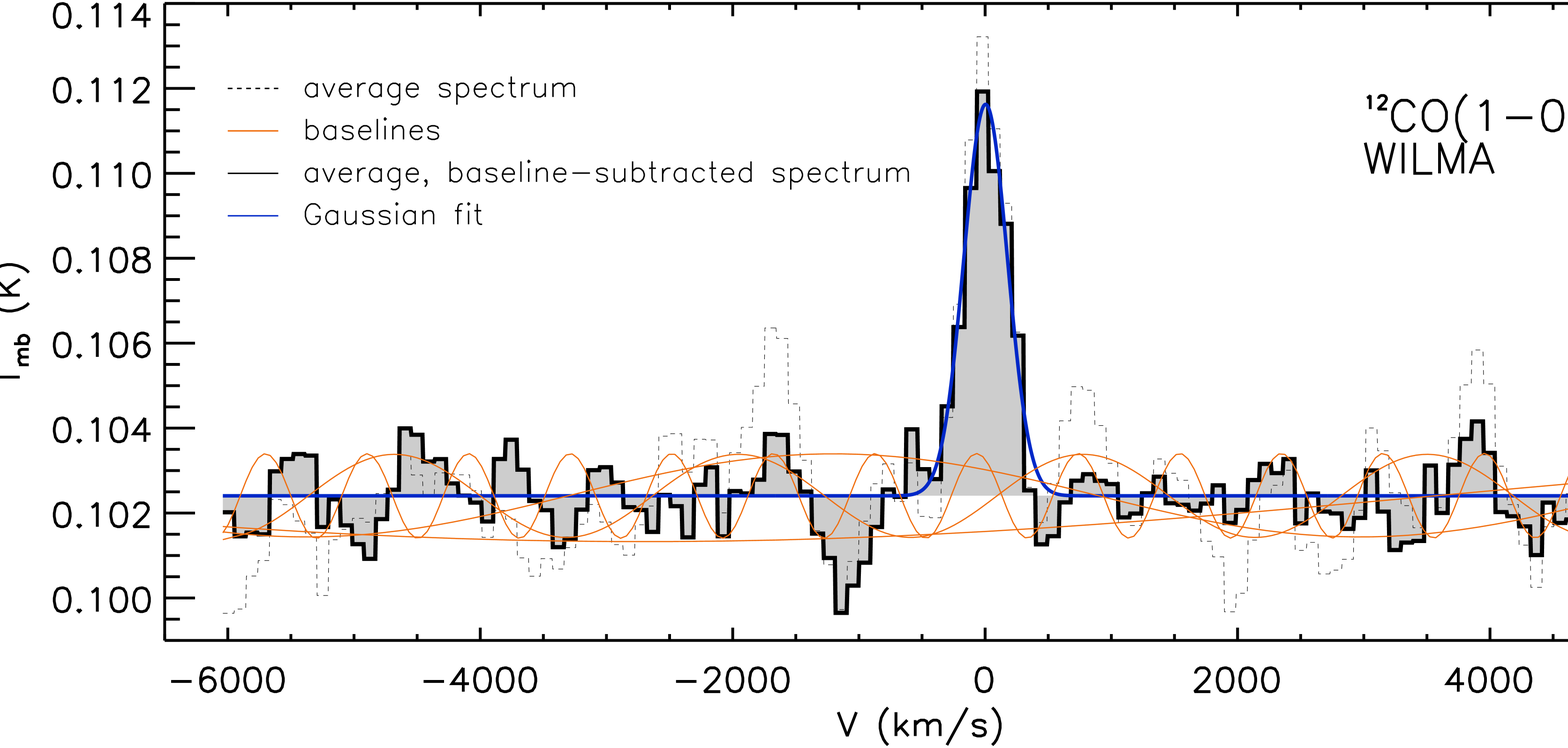} 
\includegraphics[width=5cm]{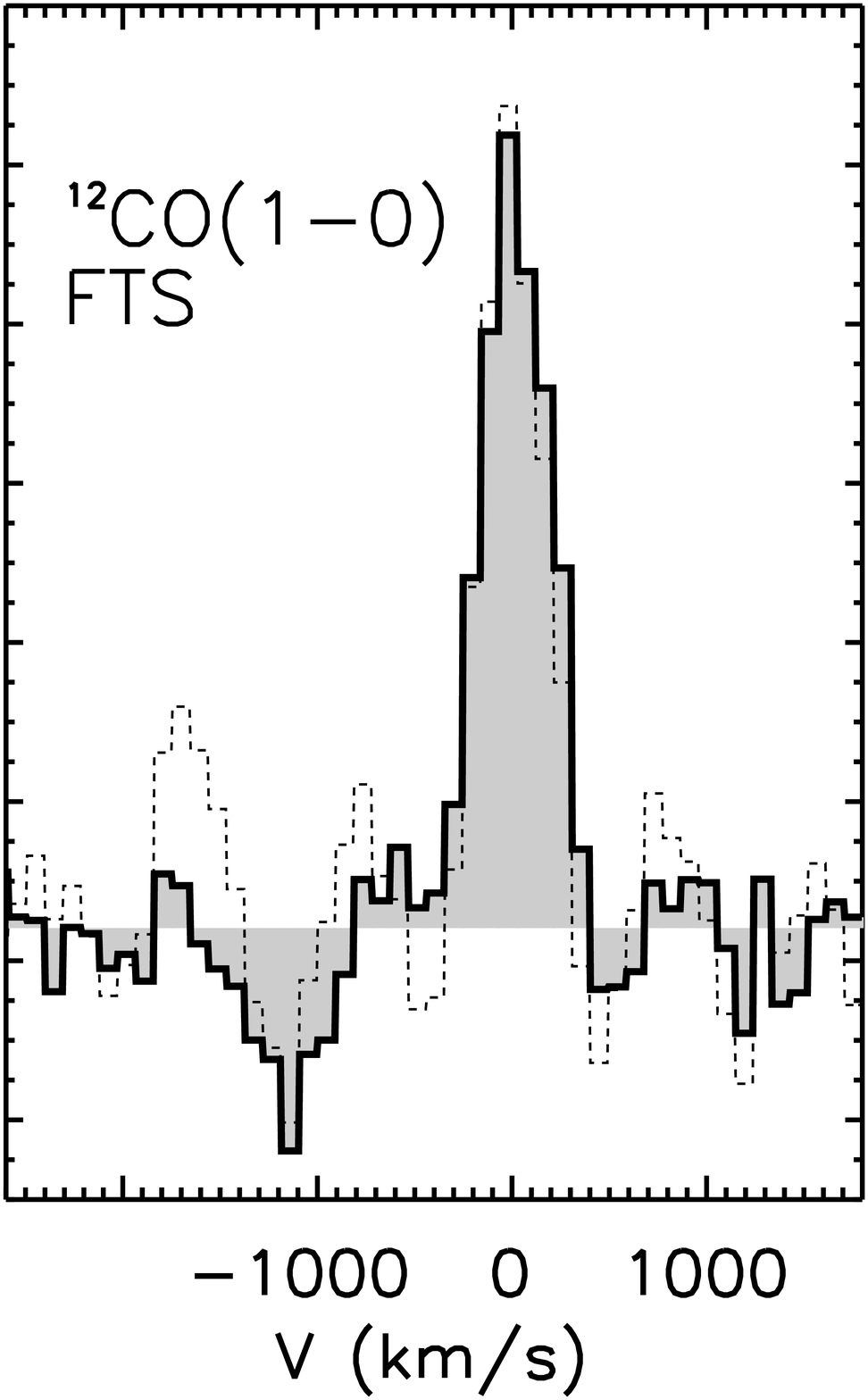} 

}
\caption{ WILMA ({\it left}) and FTS ({\it right}) \cooz\ spectrum of 4C12.50, binned to 93 \kms\ channels. The dashed histogram corresponds
to the average spectrum prior to the removal of sinusoidal baselines, which represent the harmonics of a standing wave (shown in orange). 
The final spectrum is shown as a solid, filled histogram.} 
\label{fig:CO10}
\end{center}
\end{figure*}

At 3\,mm, we averaged all useful WILMA or FTS scans, after shifting them to a reference frequency of 102.756\,GHz and removing 
bad channels from them. In contrast to the 1\,mm data, the 3\,mm data had signatures of standing waves owing to the strong synchrotron 
continuum radiation, which increases with $\lambda$. We iteratively fitted the baseline of the WILMA data with sinusoidal functions of 
periods that are integer multiples of each other, representing different harmonics of the same standing wave. The best-fit solution that 
was indicated by the CLASS baseline routine corresponded to functions with frequencies of 0.28, 0.84, 1.68, and 2.52\,GHz, and an 
amplitude of 1\,mK. These functions were removed from both the WILMA spectrum and the central part of the inner FTS spectrum. The 
final \cooz\ spectra, with a total on-source integration time of 1.2 hours, are shown in Figure~\ref{fig:CO10}. A similar approach was 
taken for the 2\,mm data. A main-beam-temperature-to-flux conversion factor of 5.0 Jy\,K$^{-1}$ was used to flux calibrate the final spectra 
at all wavelengths.


\section{Results: CO in emission and in absorption}
\label{sec:results}
The \cott\ profile of 4C12.50 consists of an emission component at the galaxy systemic velocity, i.e., at $z$$=$0.1218, and of an absorption 
feature at approximately $-$1000\kms\ from it. The CO(2$\rightarrow$3) absorption is seen in all four frequency tuning and backend combinations, 
indicating that the feature is intrinsic to the source.  Its signal-to-noise ratio (S/N) is $\sim$6 in both the WILMA and the FTS data, which have 
a corresponding root-mean-square noise of 1.7\,mK and 1.8\,mK, respectively. The outer FTS data provide further evidence that the absorption 
is real by ruling out the existence of standing waves that could create a feature of this depth. The observed optical depth at the center of 
the line, $\tau_{\rm obs,0}$, is 0.22\,($\pm$0.05), with an absorption minimum at 30\,($\pm$2)\,mJy below the local 150\,($\pm$20)\,mJy continuum. 
Averaging the best-fit Gaussian parameters of the data for each backend indicates that the minimum is at $-$950\,($\pm$90)\kms . The CO absorption 
profile remarkably resembles that of the \hi\ absorption seen in radio data \citep[Figure~\ref{fig:CO32};][]{morganti04}. Its resolution-corrected width 
is 250\,($\pm$80)\kms , reflecting the collective motions of multiple clouds along the line of sight.

An absorption line is also seen at $-$1100\kms\ away from \cooz\  after the progressive subtraction of standing wave harmonics from the 3\,mm 
baseline. However, the wave amplitude is comparable to the absorption minimum (Figure~\ref{fig:CO10}), impeding its exact profile study. 
We thus simply constrain $\tau_{\rm obs,0}$(0$\rightarrow$1) towards the 0.51 ($\pm$0.06)\,Jy continuum background to an upper limit of 0.03. 

The \cooz\ emission peaks at a main beam temperature of 9.2\,($\pm$1.2)\,mK above its underlying continuum. Its resolution-corrected 
width is 400\,($\pm$95)\kms , and its intensity, $I_{\rm {CO(1-0)}}$, is 4.0\,($\pm$1.3) K\kms (see also \citealt{evans99,evans05}).  
From this, we deduce that the cold \htwo\ gas reservoir of 4C12.50 is 1.0\,($\pm$0.3)$ \times$10$^{10}$\msun, assuming that the gas mass is 
given by the product $23.5\, \alpha\, I_{\rm {CO(1-0)}}\, \Omega_B\, {D_L^2} /{(1+z)^3} \msun $
\citep{solomon97}, where $\Omega_B$ is the telescope main beam area in arcsec$^2$, $D_L$ is the source luminosity distance in Mpc, 
and $\alpha$ is the CO luminosity to \htwo\ mass conversion factor. We used a 103\,GHz  beam of 28\arcsec\ (61\,kpc), and 
$\alpha$=0.8\msun\,(K\kms\,pc$^2$)$^{-1}$ \citep{downes98} because 4C12.50 has a ULIRG-like infrared luminosity (2.5$\times$10$^{12}$\lsun )  
based on its 12,\,25,\,60,\,and 100\um\,flux \citep{golombek88,sanders_mirabel96,guillard12}.

The \cott\ emission is detected with a S/N of 25 and 21 in the WILMA and the FTS data, respectively. Its main beam temperature peaks at 
17.5\,($\pm$1.0) mK above the local continuum. Its resolution-corrected width is 510\,($\pm$65)\kms , and its intensity is 9.9\,($\pm$1.6) K\kms . 

\begin{figure}
\begin{center}
\scalebox{1.0}{
\includegraphics[width=9cm]{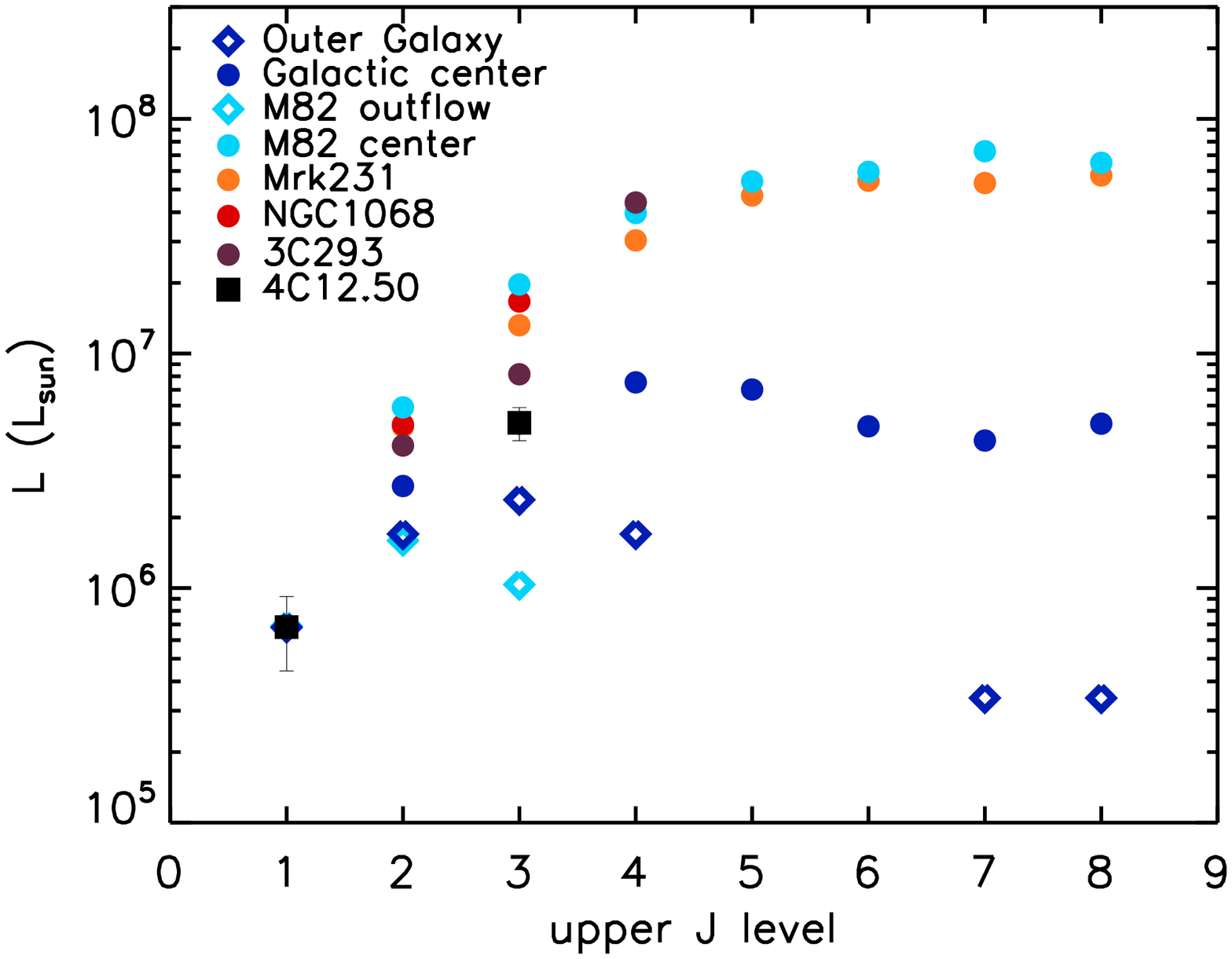} 
}
\caption{ CO line luminosities in 4C12.50 and other sources, normalized to L$_{\rm {CO(1-0)}}$(4C12.50)/L$_{\rm {CO(1-0)}}$.
The data are from \citet{fixsen99}, \citet{dumke01}, \citet{weiss01,weiss05}, \citet{papadopoulos07,papadopoulos08}, \citet{panuzzo10}, 
\citet{vanderwerf10}, and \citet{krips11}.  Typical uncertainties are 0.1-0.3 dex.
} 
\label{fig:sled}
\end{center}
\end{figure}

The $I_{\rm {CO(3-2)}}$/$I_{\rm {CO(1-0)}}$ ratio of the ambient gas, 2.5, is  below its theoretically predicted value of 9 for the case of optically thick gas 
that is thermalized to a single \tex . It is also below that of the central line of sight of M82, of several local AGN, ULIRGs, and radio galaxies, but above that 
of the outflow of M82, and similar to that of the Milky Way. This comparison is visualized in Figure~\ref{fig:sled} for the line luminosities. Even though this 
ratio is indicative of large reservoirs of cold and diffuse gas, it  does not rule out the presence of a warm and dense, highly excited CO component, which  
could be detected through high J emission lines. The non-detection of HCN (or HCO$^+$) at the 3$\sigma$ level of 0.86\,K\kms\ for a width of 460\kms\ 
cannot be used to argue for or against either interpretation. It leads to I$_{\rm CO}$/I$_{\rm HCN}$$\ge$8 \citep{gao04} for a linearly interpolated CO(2$-$1) 
intensity of 7\,K\kms .

\section{Discussion: on the properties of a multi-phase AGN-driven outflow}
\label{sec:discussion}

The rapid molecular gas motions that we detected provide direct evidence of an outflow in 4C12.50 because none of its merging components \citep{axon00} 
is moving faster than $\pm$250\kms\ from its systemic velocity \citep{holt03,zaurin07}. The maximum gas velocity reaches $-$1500\kms , which is atypical of 
supernova-driven outflows \citep[e.g.,][]{heckman00,rupke05}, but characteristic of AGN-driven outflows \citep[e.g.,][]{rupke11,sturm11}. AGN radiation 
pressure and winds can affect the gas kinematics in 4C12.50: the AGN radiation does reach and ionize part of the outflowing gas, which contains both 
Ne\,V  and O\,IV ions \citep{spoon09,dasyra11}. The gas can also be pushed by shocks created as the radio jet of 4C12.50 \citep{stanghellini97} 
propagated through the ISM. An \hi\ outflow in front of a bright radio knot \citep{morganti04} makes this scenario plausible for the neutral gas, and even for 
the molecular gas. A jet-ISM interaction can be responsible for the irregular \htwo\ kinematics \citep{dasyra_combes11,nesvadba11}, the strong \htwo\ 
emission \citep{ogle10}, and the high CO excitation \citep{papadopoulos08} of radio galaxies.

To constrain the line-of-sight CO column density, \nco, from the observed absorption, we assume that the gas is homogeneously distributed 
and in local thermodynamic equilibrium (LTE). We can then relate \nco\ to $\tau$ through
\begin{equation}
\label{eq:column}
N_{\rm CO} = 8 \pi  \frac{\nu ^3}{c^{3}} \frac{Q(T_{\rm ex})\, e^{E_{\rm J}/kT_{\rm ex} }}{g_{\rm J+1} A_{\rm J+1,J}\,(1-e^{-h\nu / kT_{\rm ex}} )} \int{\tau \, dV}
\end{equation}
\citep[e.g.,][]{wiklind95}, where $c$ is the lightspeed, $h$ is the Planck constant, $k$ is the Boltzmann constant, $E_{\rm J}$ and $g_{\rm J}$ are the 
energy and the statistical weight of the state $J$, respectively, $A_{J+1,J}$ is the Einstein coefficient for spontaneous emission, $\nu$ is the frequency of 
the emitted line, and $Q$ is the partition function at the excitation temperature \tex . The ratio $\int{\tau (2\rightarrow3)}\,dV\,/\int{\tau (0\rightarrow1)\,dV}$ 
agrees with its observed limit, $>$6.4, if \tex$\geq$65\,K. Turbulence and shocks can easily bring the gas to this \tex\ through collisions.
For \tex =65\,K, \nco\ is equal to 1.8$\times$10$^{18}$cm$^{-2}$. 

This number will increase with increasing \tex , but will remain on the same order of magnitude up to 200\,K. The filling factor of the background source 
by the molecular clouds, \fc , can play an important role in the \tex\ and \nco\ determinations. Partial coverage of the background source ($f_{\rm c}$$<$1) 
can cause the observed (beam-averaged) optical depth to deviate from its actual value, which is then equal to $-ln [1-(1-e^{-\tau_{\rm obs}})/f_{\rm c}]$ 
\citep{wiklind94}. For gas in LTE, this relation is computed from the radiative transfer equation, assuming that \tex $<<$T$_{\rm bg}$ and that the 
cosmic microwave background contribution to the gas excitation is negligible \citep[e.g.,][]{staguhn97}. Requiring the gas to be optically thick limits 
\fc\ to values above 0.2. Data of multiple transitions are needed to measure $f_c$, further constrain \nco , and examine any large velocity gradients 
\citep{scoville74} and different energy level populations \citep{israel91}.

For the present assumptions and a CO-to-H$_2$ abundance ratio of 10$^{-4}$, the \htwo\ column density towards the outflow, \nhtwo , is 
$\gtrsim$2$\times$10$^{22}$ cm$^{-2}$.  The \hi\ column density is 3$\times$10$^{21}$ cm$^{-2}$, based on both optical \nai\ data \citep{rupke05} and radio \hi\ 
data \citep[][for T$_{spin}$=1000\,K]{morganti05}. For the ambient \hi\ gas, which is located 100\,pc northwest of the radio core, the column 
density is 6$\times$10$^{20}$\,cm$^{-2}$-10$^{22}$\,cm$^{-2}$ \citep[][for T$_{spin}$=100\,K]{morganti04,curran10}. The largest uncertainty in this comparison
is the wavelength-dependent spatial distribution of the background and the foreground source. The radio core emission typically dominates the millimeter and 
even the submillimeter (850\um ; \citealt{clements10}) continuum, because the emission from diffuse jet knots is described by a steep power law that declines rapidly 
with frequency \citep{krichbaum98,krichbaum08}. We thus assume that the millimeter background source area is contained within the 2\,cm beam.

For a beam area of 3\,milliarcsec$^2$ \citep{lister03} and \nhtwo =1.8$\times$10$^{22}$cm$^{-2}$, the outflow entrains 4.2$\times$10$^3$\msun\ of cold \htwo\ gas 
in front of the millimeter core. How this mass compares with the mass of the cold \htwo\ gas that flows over all lines of sight is unknown. It therefore remains unclear 
whether the outflow carries less cold than warm H2 gas. However, the outflowing-to-ambient mass ratio of the cold gas will be lower than that of the 400\,K gas (37\%; 
\citealt{dasyra_combes11}), as long as the cold outflowing gas weighs less than 3$\times$10$^9$ \msun . If this amount of gas existed in the millimeter beam, and 
without a background source illuminating most of the clouds, the outflow would be seen in emission. Our \cozo\ data instead limit any additional outflowing cold gas 
mass to be $\le$2$\times$10$^9$\msun\,when integrating over a 1000 kpc$^2$ area and a 1500\kms\ velocity range \citep{batcheldor07}.


\section{Summary}
\label{sec:conclusions}

We have presented deep 30\,m observations of $^{12}$CO (1$-$0) and (3$-$2) in the nearby ULIRG and radio galaxy 4C12.50, which has an impressive outflow of warm (400\,K) 
\htwo\ gas \citep{dasyra_combes11}. Our observations led to the discovery of the cold component of the molecular outflow. Absorption from J$=$2$\rightarrow$3 is seen in data 
obtained with two different frequency tunings and backends. It peaks at $-$950\kms, and its profile resembles that of the \hi\ absorption component that traces the neutral 
gas outflow \citep{morganti04}. Absorption is also potentially seen for the 0$\rightarrow$1 transition. Treating the optical depth of the latter as an upper limit and assuming 
that the clouds fully cover their background source, we have found that the molecular gas has an excitation temperature of at least 65\,K in LTE conditions. For this temperature, 
\nhtwo\ is 1.8$\times$10$^{22}$cm$^{-2}$. This suggests that the outflow carries 4.2$\times$10$^3$\msun\ of cold \htwo\ gas along the nuclear line of sight. Even if the mass 
of the cold outflowing gas could exceed that of the warm outflowing gas, 5.2$\times$10$^7$\msun , for a higher \nco\ value or when integrating over several lines of sight, it 
would most likely not exceed 3$\times$10$^{9}$\msun , i.e., one third of the ambient cold gas reservoir. The outflowing-to-ambient mass ratio could thus be elevated into the 
warm gas phase with respect to the cold gas phase.

\begin{acknowledgements}
 K. D. acknowledges support by the Centre National d\'\,Etudes Spatiales (CNES), and is  thankful to P. Salom\'e for technical tips and to P. Papadopoulos for 
 useful discussions. Based on data obtained with the 30\,m telescope of IRAM, which is supported by INSU/CNRS (France), MPG (Germany), and IGN (Spain).
\end{acknowledgements}


{}

\end{document}